\documentclass[global,twocolumn]{svjour}
\usepackage{graphicx}
\usepackage{color}

\journalname{Applied Physics B}
\begin{document}

\title{Dynamical behaviour of birefringent Fabry-Perot cavities}

\author{Paul Berceau\inst{1}, Mathilde Fouch\'e\inst{2,3}, R\'emy Battesti\inst{1}, Franck Bielsa\inst{1}, Julien Mauchain\inst{1} and Carlo Rizzo\inst{2,3}}

\institute{Laboratoire National des Champs Magn\'etiques Intenses
(UPR 3228, CNRS-INSA-UJF-UPS), F-31400 Toulouse Cedex, France \and Universit\'e
de Toulouse, UPS, Laboratoire Collisions Agr\'egats R\'eactivit\'e,
IRSAMC, F-31062 Toulouse, France \and CNRS, UMR 5589, F-31062
Toulouse, France}

\offprints{mathilde.fouche@irsamc.ups-tlse.fr}

\maketitle

\begin{abstract}
In this paper we present a theoretical and experimental study of the
dynamical behaviour of birefringent cavities. Our experimental data
show that usual hypothesis which provides that a Fabry-Perot cavity
is a first-order low pass filter cannot explain the behaviour of a
birefringent cavity. We explain this phenomenon, and give the
theoretical expression of the equivalent cavity filter which
corresponds to a second-order low pass filter.
\end{abstract}


\section{Introduction}

Fabry-Perot cavities are widely used in experiments devoted to the
detection of very small optical effects, \textit{e.g.} in the framework of
gravitational wave interferometers \cite{WaveDetector},
optomechanical noise studies \cite{Heidmann}, frequency measurements
via optical clocks \cite{OpticalClock}, Lorentz invariance
experimental tests \cite{Lorentz}, or vacuum magnetic birefringence
measurements \cite{BMV,epjdequipe}.

Fabry-Perot cavities made with interferential mirrors are
birefringent
\cite{BirMirror,BirMirror_Lee,BirMirror_Morville,BirMirror_Huang}.
For most of the Fabry-Perot fundamental applications, this property
can be neglected, at least at first sight, since the studied effects
do not depend on polarization. Obviously, this is not the case of
birefringence studies reported in refs.\,\cite{BMV,epjdequipe}.

The dynamical behaviour of non birefringent cavities has been
studied in details \cite{Uehara95}. The cavity acts as a first-order
low pass filter whatever the polarization of the incident light is,
and the frequency spectrum of the transmitted light is modified
consequently. As far as we know, nothing has been published so far
regarding birefringent cavities. In this paper we present a
theoretical and experimental study of the dynamical behaviour of
birefringent cavities in the presence of a time variation of the
incident light intensity and in the presence of a time variation of
the birefringence itself.

Our experimental data show that a birefringent cavity cannot be
described as a first-order low pass filter as it is generally
assumed. We explain this phenomenon, and give the theoretical
expression of the equivalent cavity filter which corresponds to a
second-order low pass filter. We also discuss the implications of
this cavity behaviour in the case of existing experiments for
measuring very low birefringence effects using Fabry-Perot cavities.

\section{Experimental Setup}
\label{Sec:Setup}

\begin{figure}[h]
\begin{center}
\resizebox{1\columnwidth}{!}{
\includegraphics{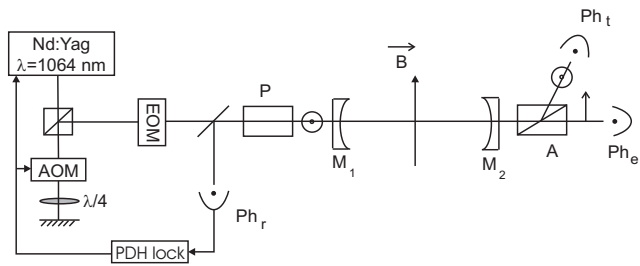}}
\caption{\label{Fig:ExpSetup} Experimental setup. A Nd-YAG laser is
frequency locked to the Fabry-Perot cavity made of mirrors M$_1$ and
M$_2$. The laser beam is linearly polarized by the polarizer P and
analyzed with the polarizer A. This analyzer allows to extract the
extraordinary beam sent on photodiode Ph$_\mathrm{e}$ as well as the
ordinary beam sent on photodiode Ph$_\mathrm{t}$. The beam reflected
by the cavity analyzed on the photodiode Ph$_\mathrm{r}$ is used for the cavity locking. A
transverse magnetic field $B$ can be applied inside the cavity in
order to study the magnetic birefringence of the medium. EOM =
electro-optic modulator; AOM = acousto-optic modulator.}
\end{center}
\end{figure}

Our study is performed in the framework of the BMV experiment
\cite{epjdequipe} whose goal is to measure vacuum magnetic
birefringence. Briefly, as shown on Fig.\,\ref{Fig:ExpSetup}, a
linearly polarized Nd:Yag laser beam ($\lambda = 1064$\,nm) is
injected into a Fabry-Perot cavity made of mirrors M$_1$ and M$_2$.
The length of the cavity is $L=2.2$\,m. The laser frequency is
locked to the cavity resonance frequency using the Pound-Drever-Hall
method \cite{PDH}. To this end, the laser is phase-modulated at 10\,MHz with an electro-optic modulator (EOM). The
beam reflected by the cavity is then analyzed on the photodiode
Ph$_\mathrm{r}$. This signal is used to drive the acousto-optic
modulator (AOM) frequency for a fast control and the Peltier element
of the laser for a slow control.

Our birefringence measurement is based on an ellipticity
measurement. Light is polarized just before the cavity by the
polarizer P. The beam transmitted by the cavity is then analyzed by
the analyzer A crossed at maximum extinction and collected by a low
noise photodiode Ph$_\mathrm{e}$. The analyzer has an escape window
which allows us to extract the reflected ordinary beam. This beam is
collected by the photodiode Ph$_\mathrm{t}$. Both signals are
simultaneously used in the data analysis as following:
$I_\mathrm{e}/I_\mathrm{t} = \sigma^2 + \Psi_\mathrm{tot}^2$, where
$\Psi_\mathrm{tot}$ is the total ellipticity acquired by the beam
going from P to A and $\sigma^2$ is the polarizer extinction ratio.
Our polarizers are Glan Laser Prism manufactured by Karl Lambrecht
Corporation (Chicago, USA) which have an extinction ratio of
$4\times 10^{-7}$.

The origin of the total ellipticity cavity is firstly due to the
mirror intrinsic birefringence. Mirrors are similar to wave plates. For small birefringence, combination of
both wave plates gives a single wave plate. The phase retardation
and the axis orientation of this equivalent wave plate depends on the birefringence of each mirror and on their respective orientation
\cite{Jacob,brandi}. We define the ellipticity induced on the
linearly polarized laser beam by the Fabry-Perot cavity as $\Gamma$
which is set to about $10^{-2}$ in the experiment described in this
paper.

A second component of the total ellipticity appears when a
birefringent medium is placed inside the cavity. For example, on
magnetic birefringence measurements, a transverse magnetic field $B$
is applied inducing an ellipticity $\Psi\propto B^2l$ where $l$ is
the optical path in the magnetic field.

Finally, if ellipticities are small compared with unity, one gets:
\begin{equation}
I_\mathrm{e}/I_\mathrm{t} = \sigma^2 + (\Gamma + \Psi)^2. \label{Eq:Ellipticity}
\end{equation}

The goal of the experiment presented in this paper is to have a
complete understanding of birefringent cavity dynamical behaviour.
For this study, two different methods have been implemented. In the
next section we present the cavity behaviour in the case of a time variation of
the incident light intensity whereas in the last section, the
ellipticity inside the cavity is modulated.

\section{Time variation of the incident light intensity}

In this part, we study the cavity dynamical behaviour to a time
variation of the incident laser beam intensity while the total
ellipticity remains constant. Two approaches have been used: study
of the cavity response to a step function or to an intensity
frequency modulation of the incident beam. The first section is
devoted to the presentation of both approaches when looking at the
ordinary beam collected by Ph$_\mathrm{t}$ $i.e.$ when the
transmitted beam polarization is parallel to the incident one. In the second section, this study is performed on the
extraordinary beam $i.e.$ when the beam polarization is
perpendicular to the incident one.

\subsection{Cavity dynamical behaviour towards the ordinary beam}
\label{SubSec:I_ord}

\subsubsection{Time response of the cavity to a step function\\}

The simplest way to study the cavity response is to abruptly switch
off the intensity of the incident beam locked to the cavity and then
to look at the intensity decay of the beam transmitted by the
cavity. This method allows to determine typical cavity parameters as
the photon lifetime, the cavity finesse, the full width at half
maximum or the cavity quality factor.

Experimentally, the intensity is switched off thanks to the
acousto-optic modulator (AOM) shown on Fig.\,\ref{Fig:ExpSetup} and
used as an ultrafast commutator. Its switched-off time is less than
$1\,\mu$s, far less than the photon lifetime as we will see below.
On Fig.\,\ref{Fig:It_t} the intensity of the ordinary beam is
plotted as a function of time. For $t<t_0$, the laser is locked to
the cavity. The laser intensity is switched off at $t_0$. For
$t>t_0$, one sees the typical exponential decay \cite{svelto}:
\begin{eqnarray}
I_\mathrm{t}(t)=I_\mathrm{t}(t_{0})e^{-(t-t_{0})/\tau},\label{Eq:DureeVie_Iord}
\end{eqnarray}
where $\tau$ is the photon lifetime. This lifetime is related to the
finesse $F\simeq \pi/(1-R)$ of the cavity through the relation: $\tau =
LF/\pi c$ with $c$ the speed of light and $R$ the mirror
reflectivity which is supposed to be the same for both mirrors. By
fitting our data with this expression one gets $\tau =(245\pm
10)\,\mu$s corresponding to a finesse of $F=(105\pm 5) \times 10^3$. The uncertainty results from statistical uncertainty.

\begin{figure}[h]
\begin{center}
\resizebox{1\columnwidth}{!}{
\includegraphics{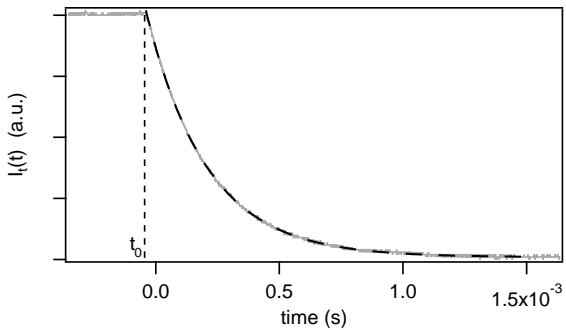}}
\caption{\label{Fig:It_t} Time evolution of the intensity of the
ordinary beam (gray line). The laser is switched off at $t=t_{0}$.
Experimental data are fitted by an exponential decay
(black dashed line) giving a photon lifetime of $\tau =(245\pm
10)\,\mu$s and a finesse of $F=(105\pm 5) \times 10^3$.}
\end{center}
\end{figure}

\subsubsection{Frequency response of the cavity to an intensity modulation\\}

In order to complete our understanding of the experiment, we also
study the frequency response of the Fabry-Perot cavity to an
intensity modulation. Theoretically, for an incident light modulated
in intensity at pulsation $\omega_\mathrm{F}$ and for a small depth
of modulation, the complex response function is given by
\cite{Uehara95}:
\begin{eqnarray}
H_\mathrm{t}\left( \omega_\mathrm{F}\right) &=&
\frac{I_\mathrm{t}^{(\omega_\mathrm{F})}}{I_\mathrm{i}^{(\omega_\mathrm{F})}}
\propto \frac {1}
{1+i\frac{\omega_\mathrm{F}}{\omega_\mathrm{c}}}.\label{Eq:Ht}
\label{Eq:transfunc}
\end{eqnarray}
$I_\mathrm{t}^{(\omega_\mathrm{F})}$
($I_\mathrm{i}^{(\omega_\mathrm{F})}$) is the $\omega_\mathrm{F}$ component of the ordinary (incident) beam intensity. The
response function operates as a first-order low pass filter with a
cutoff frequency $\nu_\mathrm{c} = \omega_\mathrm{c}/2\pi =
1/4\pi\tau$.

Experimentally, to study the cavity frequency response, the laser is
locked to the cavity and the intensity is modulated with a small
depth of modulation thanks to the AOM. The intensity of the incident
beam and of the ordinary beam transmitted by the cavity is recorded
at different modulation frequencies.

\begin{figure}[h]
\begin{center}
\resizebox{1\columnwidth}{!}{
\includegraphics{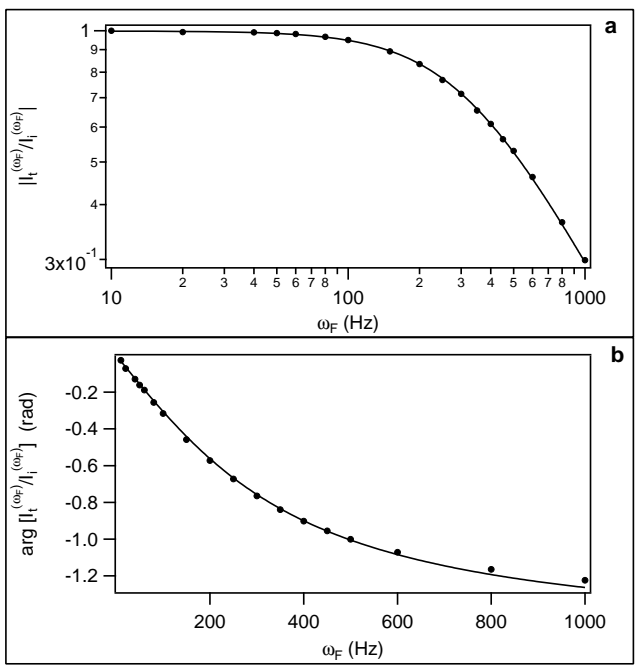}
} \caption{\label{Fig:Iord_Irefl} Experimental cavity response
function towards the ordinary beam. (a) Gain of the response
function normalized to 1 at low frequency $i.e.$
$|I_\mathrm{t}^{(\omega_\mathrm{F})}/I_\mathrm{i}^{(\omega_\mathrm{F})}|$
as a function of the modulation frequency $\omega_\mathrm{F}$. Data
are fitted by the gain of a first-order low pass filter. (b) Phase
delay between $I_\mathrm{t}^{(\omega_\mathrm{F})}$ and
$I_\mathrm{i}^{(\omega_\mathrm{F})}$ as a function of the modulation
frequency. Data are fitted by the phase delay of a first-order low
pass filter.}
\end{center}
\end{figure}

Results are presented on Fig.\,\ref{Fig:Iord_Irefl}.
Fig.\ref{Fig:Iord_Irefl}a presents the gain of the response function normalized to 1 at low
frequency and Fig.\ref{Fig:Iord_Irefl}b presents the phase delay.
Data are fitted by the response function of a first-order low pass
filter. Cutoff frequency is equal to $\nu_\mathrm{c}=(310\pm20)$\,Hz
when fitting the gain, and $\nu_\mathrm{c}=(315\pm20)$\,Hz when
fitting the phase delay. These values correspond to a finesse of
respectively $F=(109\pm9) \times 10^3$ and $F=(108\pm8) \times
10^3$, which is in agreement with the finesse measured with
the previous approach.
\\

While in the second approach we are looking at the frequency response of the cavity, the first approach is performed in the time domain. Both areas of analysis are equivalent and can be connected thanks to Laplace transform. However, the time analysis is usually preferred to the frequency analysis since it is simpler and quicker to implement on the experiment.

Finally, the study performed on the ordinary beam shows that the
dynamical behaviour of our cavity is the same as the one obtained on
non birefringent cavities. The typical exponential decay is observed
when the incident light is suddenly switched off and the frequency
response shows that the cavity behaves as a first-order low pass
filter.

\subsection{Cavity dynamical behaviour towards the extraordinary beam}

We now turn to the study on the extraordinary beam collected by
Ph$_\mathrm{e}$ $i.e.$ the beam transmitted by the cavity with a
polarization perpendicular to the polarization of the incident one.

\subsubsection{Time response of the cavity to a step function\\}

\begin{figure}[h]
\begin{center}
\resizebox{1\columnwidth}{!}{
\includegraphics{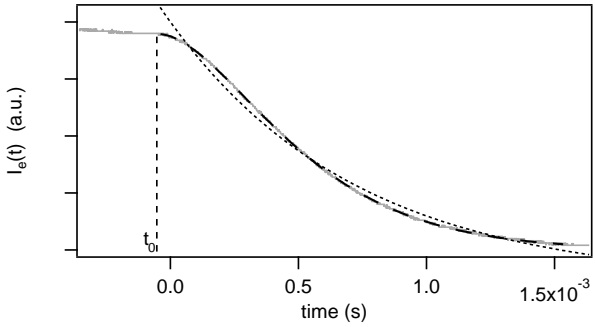}
}\caption{\label{Fig:Iext_t} Time evolution of the intensity of the
extraordinary beam (gray line). The laser is switched off at
$t=t_{0}$. Experimental data are fitted by
Eq.\,(\ref{Eq:DurreVie_Iext}) (black dashed line) giving a photon
lifetime of $\tau = (245\pm10)\,\mu$s. The fit with an exponential decay (dots) does not correspond to the experimental behaviour and gives a photon lifetime of $\tau = 735\,\mu$s in disagreement with previously given values.}
\end{center}
\end{figure}

Time evolution of the extraordinary beam when the incident beam is
suddenly switched off is shown on Fig.\,\ref{Fig:Iext_t}. By
comparing this curve to the one plotted on Fig.\,\ref{Fig:It_t}, we
see that the cavity does not have the same behaviour for
$I_\mathrm{t}$ and $I_\mathrm{e}$. When one fits $I_\mathrm{e}$ with an exponential decay, the experimental behaviour is not reproduced and it gives a photon lifetime of $\tau = 735\,\mu$s in disagreement with previously given values. We will show that this is due to the intrinsic birefringence of the cavity.

Let's calculate the transmitted intensity along the round-trip
inside the cavity:
\begin{itemize}
\item For $t \leq t_{0}$, the laser is continuously locked to the
cavity. According to Eq.\,(\ref{Eq:Ellipticity}), the intensities of the ordinary and the extraordinary beams
are related by:
\begin{eqnarray*}
I_\mathrm{e}(t\leq t_{0})=\Gamma^{2}I_\mathrm{t}(t\leq t_{0}).
\end{eqnarray*}
The polarizer extinction ratio is neglected since we have
$\sigma^2\ll \Gamma^2$ and no birefringence is applied inside the
cavity.

\item At $t = t_{0}$, the laser beam is abruptly switched off, the cavity
empties gradually. The ordinary and extraordinary beams are slightly
transmitted at each reflection on the mirrors. But, because these
mirrors are birefringent, some photons of the ordinary beam are
converted into the extraordinary one. The reverse effect is
neglected because $I_\mathrm{e}\ll I_\mathrm{t}$.

As shown on Eq.\,(\ref{Eq:Ellipticity}), the total ellipticity corresponds to the sum of ellipticities when they are small. Furthermore, following Ref.\,\cite{brandi}, the ellipticity $\Gamma$ induced by the cavity is related to the ellipticity induced per round-trip $\gamma$ trough the relation: $\gamma=\Gamma\pi/F$.

Thus after one round-trip inside the cavity, $i.e.$ at time $t_0 + t_\mathrm{D} =
t_0+ 2L/c$, we get:
\begin{eqnarray*}
I_\mathrm{e}(t_{0}+t_\mathrm{D})=\left( \Gamma + \gamma\right)^{2}
I_\mathrm{t}(t_{0}+t_\mathrm{D}).
\end{eqnarray*}

\item After $p$ round-trips, one gets the intensity of the extinction beam:
\begin{eqnarray}
I_\mathrm{e}(t_{0}+pt_\mathrm{D})=\left( \Gamma + p\gamma
\right)^{2} I_\mathrm{t}(t_{0}+pt_\mathrm{D}).\label{Eq:Ie_p}
\end{eqnarray}
\end{itemize}
Assuming that Eq.\,(\ref{Eq:Ie_p}) holds not only at times
$t_0+pt_\mathrm{D}$ but also at any time $t>t_0$ and using Eq.\,(\ref{Eq:DureeVie_Iord}) for $I_t$, we can write:
\begin{eqnarray}
I_{e}(t)=I_{e}(t_{0})\left(1 + \frac{t-t_{0}}{2\tau} \right)^{2}
e^{-\frac{t-t_{0}}{\tau}}. \label{Eq:DurreVie_Iext}
\end{eqnarray}
This expression is used to fit our experimental data plotted on
Fig.\,\ref{Fig:Iext_t}. We find a photon lifetime of $\tau = (245
\pm 10)\,\mu$s which is in good agreement with the value found in the previous section.

\subsubsection{Frequency response of the cavity to an intensity modulation\\}

\begin{figure}[h]
\begin{center}
\resizebox{1\columnwidth}{!}{
\includegraphics{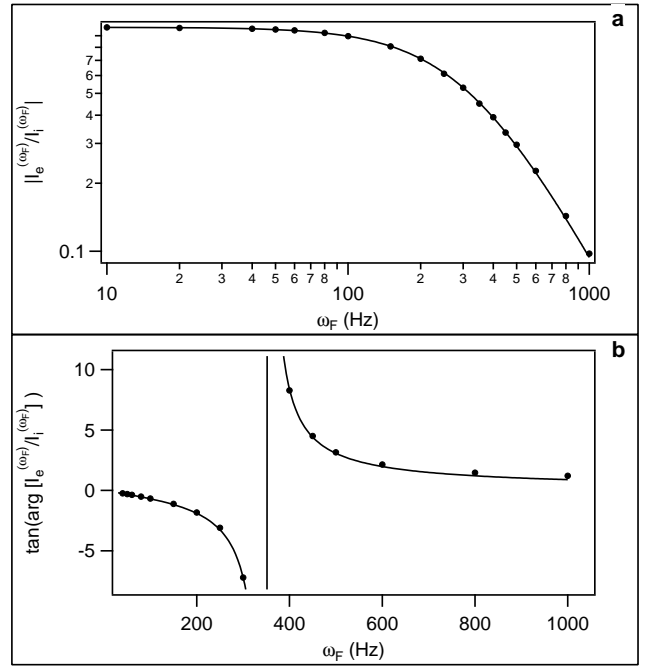}
}\caption{\label{Fig:Iext_Irefl} Cavity response function towards
the extraordinary beam. (a) Gain of the response function normalized
to 1 at low frequency $i.e.$
$|I_\mathrm{e}^{(\omega_\mathrm{F})}/I_\mathrm{i}^{(\omega_\mathrm{F})}|$
as a function of the modulation frequency $\omega_\mathrm{F}$. Data
are fitted by the gain of a second-order low pass filter. (b) Tangent of the phase
delay between $I_\mathrm{e}^{(\omega_\mathrm{F})}$ and
$I_\mathrm{i}^{(\omega_\mathrm{F})}$ as a function of the modulation
frequency. Data are fitted by the phase delay of a second-order low
pass filter.}
\end{center}
\end{figure}

As done before, we also study the frequency response of the cavity
towards the extraordinary beam to an intensity modulation. Results
are presented on Fig.\,\ref{Fig:Iext_Irefl}.

To calculate the complex response function expected theoretically,
we use Eq.\,(\ref{Eq:DurreVie_Iext}) and the Laplace transform and we
get:
\begin{eqnarray*}
H_\mathrm{e}\left( \omega_\mathrm{F}\right) &=&
\frac{I_\mathrm{e}^{(\omega_\mathrm{F})}}{I_\mathrm{i}^{(\omega_\mathrm{F})}}
\propto \left( \frac {1}
{1+i\frac{\omega_\mathrm{F}}{\omega_\mathrm{c}}} \right)^{2}.
\label{Eq:transfunc}
\end{eqnarray*}
$I_\mathrm{e}^{(\omega_\mathrm{F})}$ corresponds to the $\omega_\mathrm{F}$ component of the extraordinary beam intensity. The
response function operates as a second-order low pass filter with
the same cutoff frequency $\nu_\mathrm{c}$ found previously for the
ordinary beam. Data of Fig.\,\ref{Fig:Iext_Irefl} are fitted by the following expressions:
\begin{eqnarray}
|H_\mathrm{e,n}\left( \omega_\mathrm{F}\right)| = \frac{1}{1+\left(
\frac{\omega_\mathrm{F}}{\omega_\mathrm{c}}\right)^2}\\
\arg[H_\mathrm{e,n}\left( \omega_\mathrm{F}\right)] =
-\frac{2\frac{\omega_\mathrm{F}}{\omega_\mathrm{c}}}{1-\left(
\frac{\omega_\mathrm{F}}{\omega_\mathrm{c}}\right) ^{2}}.
\end{eqnarray}
Cutoff frequencies given by the fits are $\nu_{c}=(325\pm20)$\,Hz
and $\nu_{c}=(350\pm20)$\,Hz and are consistent with the values
found in the previous section.
\\

The study presented in this part shows that a birefringent cavity
cannot be described as a first-order low pass filter as it is
generally assumed for usual cavities. For the extraordinary beam,
the cavity acts as a second-order low pass filter instead of a
first-order. This filter represents the combined action of two
successive first-order low pass filters. While the first filter
characterizes the usual cavity behaviour as seen in section
\ref{SubSec:I_ord}, we can interpret the second filter in terms of
pumping or filling: due to the mirror birefringence, some photons of
the ordinary beam are gradually converted into the extraordinary
beam at each reflection.

\section{Time variation of the birefringence}

The second method implemented to study the cavity dynamical
behaviour consists in varying the cavity birefringence itself. The
intrinsic cavity birefringence can hardly be modulated. We have
chosen to obtain a time variation of the cavity birefringence by a
variation of the birefringence of the medium placed inside the
cavity.

According to Eq.\,(\ref{Eq:Ellipticity}), the measured signal is given by:
\begin{eqnarray*}
I_\mathrm{e}(t)/I_\mathrm{t} = \sigma^2 + \Gamma^2 + 2\Gamma\Psi(t).
\end{eqnarray*}
We assume that $\Psi\ll\Gamma$. Let's consider that the ellipticity
per round-trip $\psi$ applied inside the cavity is modulated with a
pulsation $\omega_\mathrm{F}$:
\begin{eqnarray*}
\psi(t) = \psi_0\sin(\omega_\mathrm{F} t).
\end{eqnarray*}
Following calculations performed in \cite{epjdequipe},
the ellipticity outside of the cavity induced by the applied
birefringence is:
\begin{eqnarray}
\Psi(t) =
\frac{\Psi_0}{\sqrt{1+\big(\frac{\omega_\mathrm{F}}{\omega_\mathrm{c}}\big)^2}}
\sin(\omega_\mathrm{F} t+\phi) \label{Eq:B_filter}
\end{eqnarray}
with $\tan \phi = -\omega_\mathrm{F}/\omega_\mathrm{c}$ and $\Psi_0 = \psi_0 F/\pi$. We see that
this ellipticity corresponds to an ellipticity filtered by a
first-order low pass filter with a cutoff frequency corresponding to
the one of the cavity. In other words, if the ellipticity $\psi$
varies over the photon lifetime in the cavity, the ellipticity
outside of the cavity is attenuated and does not remain in phase
with $\psi$.

From the experimental point of view, the birefringence inside the
cavity corresponds to a magnetic birefringence. The induced
ellipticity per round-trip is given by: $\psi \propto B^2
\sin2\theta$ where $\theta$ is the angle between light polarization
and the direction of the transverse magnetic field. To modulate this
ellipticity, one can modulate the value of the magnetic field or
modulate the direction of the magnetic field.

\begin{figure}[h]
\begin{center}
\resizebox{1\columnwidth}{!}{
\includegraphics{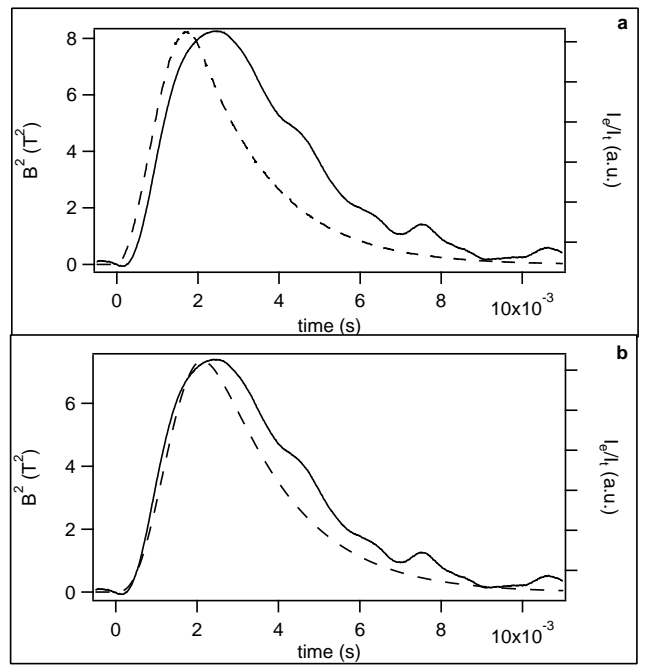}
}\caption{\label{Fig:Psi_Bfilter} (a) Dashed curve: Square of the magnetic
field as a function of time. Line: Signal $I_\mathrm{e}/I_\mathrm{t}$ as a function of time while the laser is locked to the cavity. (b) Dashed curve: Square of the magnetic
field filtered by a first-order low pass filter corresponding to the cavity filtering. Line: Signal $I_\mathrm{e}/I_\mathrm{t}$ as a function of time while the laser is locked to the cavity. Shift of
both maxima are compensated when the cavity filtering is taken into account. Noise observed on the transmitted intensities after 2\, ms of magnetic pulse are due to vibrations induced on the cavity by the magnetic pulse. This part is not taken into account in the data analysis.}
\end{center}
\end{figure}

On our experiment, the magnetic field is created thanks to pulsed
coils. Thus, the time variation of the applied birefringence
corresponds to a time variation of the square of the magnetic field.
On Fig.\,\ref{Fig:Psi_Bfilter}a, a typical magnetic pulse is plotted. It reaches its maximum of 2.9\,T within less than 2\,ms.

The cavity finesse is 100000 which corresponds to a photon lifetime of 230\,ms. About 15\,mbar of air was inserted inside the vacuum
chamber which contains the cavity and the polarizers. The applied birefringence is always smaller compared to the mirror birefringence. The observed signal is shown on Fig.\,\ref{Fig:Psi_Bfilter}a and b on the right axis and compared to the magnetic field. We see that both maxima of $B^2$ and $I_\mathrm{e}/I_\mathrm{t}$ do not coincide. But as expected by Eq.\,(\ref{Eq:B_filter}) and shown on Fig.\,\ref{Fig:Psi_Bfilter}b, this shift is actually compensated if we apply a first-order low pass filter corresponding to the cavity filtering on the square of the magnetic field.

Finally, the value of the magnetic birefringence is calculated
through the correlation between $\Psi(t)$ and $B^2(t)$ filtered
\cite{epjdequipe}. In the case of Fig.\,\ref{Fig:Psi_Bfilter} this analysis is not performed for $t>2$\,ms where vibrations are induced on the cavity due to the magnetic pulse. Improvements are currently under development to minimize this effect. If the filter is not applied on the magnetic field \textit{i.e.} if the cavity influence is not taken into account, a systematic uncertainty of a few percents is added on the value of the magnetic birefringence.

\section{Conclusion}

We have studied the dynamical behaviour of birefringent Fabry-Perot
cavities. Actually, because of the intrinsic mirror birefringence
all Fabry-Perot cavities are birefringent, and our study applies to
all of them. We have shown that the cavity dynamical behaviour
depends on polarization.

For intensity modulation of the incoming beam, its frequency
spectrum is filtered by the cavity differently depending on the
polarization of the light exiting the cavity. This filtering also
applies to the intensity noise frequency spectrum.

We have also considered the case of a cavity birefringence time
variation. To study how a cavity filters such a modulation, we have
measured a magnetic birefringence induced by a pulsed magnetic field
on a medium inside a Fabry-Perot cavity. We have experimentally
shown that depending on the photon lifetime in the cavity \textit{i.e.}
the cavity cutoff frequency, the induced ellipticity is attenuated
and becomes out of phase with respect to the magnetic field pulse.
The finesse of the cavity we used is of the order of 100000. A
higher finesse will correspond to a more important filtering and to
a bigger systematic uncertainty correction.

The problem is exactly the same if the value of the magnetic field
remains fixed while its direction compared to the cavity
birefringence axis is rotated as it is the case on other experiments
measuring magnetic birefringence. For example, in
ref.\cite{PVLAS2009} where the Cotton-Mouton effect in helium is
measured, a superconducting dipole magnet rotating at a frequency of
$0.35\,Hz$ is used. The finesse is $100000$ corresponding to a
cavity cutoff frequency of $\nu_\mathrm{c} = 116.5$\,Hz. Taking into
account the cavity filtering allows to avoid a systematic
uncertainty of $1.8\times10^{-3}\,\%$ on the final magnetic
birefringence. In the same way, in ref. \cite{QA2009}, where the
Cotton-Mouton effect of different gases is measured, a dipole
permanent magnet is rotating at about 6.8\,Hz inside a cavity with a
cutoff frequency of $725$\,Hz. The systematic uncertainty is then
$1.7\times10^{-2}\,\%$. Systematic uncertainty on such experiments
is negligible compared to statistical uncertainties, but it will
become more important if the rotating frequency increases and/or the
cavity finesse increases.

\section{Acknowledgements}
This work has been performed in the framework of the BMV project. We
thanks all the members of the BMV collaboration, and in particular
Hugo Bitard, G. Bailly and M. Nardone. We acknowledge the support of
the {\it ANR-Programme non th\'{e}matique} (ANR-BLAN06-3-139634),
and of the {\it CNRS-Programme National Particule Univers}.

\end{document}